%% file: main.tex
\documentclass[letterpaper,twocolumn,10pt]{article}
\usepackage{usenix2023_SOUPS}
\usepackage{tikz}
\usepackage{amsmath}
\usepackage{booktabs}

\begin{document}

\date{}

\title{\Large \bf Recruiting Teenage Participants for an Online Security Experiment:\\ A Case Study Using Peachjar}

\def\plainauthor{Elijah Bouma-Sims, Lily Klucinec, Mandy Lanyon, Lorrie Cranor, and Julie Downs}

\author{
{\rm Elijah Bouma-Sims}\\
Carnegie Mellon University
\and
{\rm Lily Klucinec}\\
Carnegie Mellon University
\and
 {\rm Mandy Lanyon}\\
Carnegie Mellon University
\and
 {\rm Lorrie Faith Cranor}\\
Carnegie Mellon University
\and
 {\rm Julie Downs}\\
Carnegie Mellon University
} 

\maketitle
\thecopyright

\begin{abstract}
The recruitment of teenagers for usable privacy and security research is challenging, but essential. This case study presents our experience using the online flier distribution service Peachjar to recruit minor teenagers for an online security experiment. By distributing fliers to 90 K-12 schools, we recruited a diverse sample of 55 participants at an estimated cost per participant of $\$43.18$. We discuss the benefits and drawbacks of Peachjar, concluding that it can facilitate the recruitment of a geographically diverse sample of teens for online studies, but it requires careful design to protect against spam and may be more expensive than other online methods. We conclude by proposing ways of using Peachjar more effectively.
\end{abstract}

\input{secs/01_intro}
\input{secs/02_methods}
\input{secs/03_results}

\input{secs/04_discussion}

\section*{Acknowledgments}
This study was partially funded by a Carnegie Mellon University CyLab Presidential Fellowship and the Innovators Network Foundation.

\bibliographystyle{plain}
\bibliography{main}

\input{secs/05_app}

\end{document}

%% file: secs/01_intro.tex
\section{Introduction}

The recruitment of minor teenagers for usable security and privacy research is challenging, but important. Accessing a sample often requires researchers to work with local schools or other ``gatekeepers'' who can facilitate access to participants~\cite{horowicz2023recruiting}. Although these gatekeepers can provide valuable insight into the needs of their local community, researchers without existing connections may find it difficult to begin working with minors. In this context, Peachjar\footnote{\url{https://www.peachjar.com/}} is an intriguing solution for expanding the research recruitment of minors. Peachjar is a digital flier management system for K-12 schools that allows third parties to pay to submit fliers that, if approved, are emailed to parents in the selected school district and posted on a website. This service enables researchers to directly reach parents in a school district, avoiding the labor-intensive process of establishing relationships with local schools. Moreover, with 780 participating school districts in 42 US states as of May 16, 2024, Peachjar could allow researchers to reach a more geographically diverse sample than they might otherwise reach.
Furthermore, Peachjar allows researchers to directly target parents of school-aged children.

Peachjar has previously been used to recruit minor participants across age groups, especially in medical research~\cite{johnstone2022micro,cohen2021effect,cohen2021impact}. These studies seem to have targeted schools local to the researchers and have  recruited only part of their sample from Peachjar. For example, Cohen et al.~\cite{cohen2021effect} used a combination of techniques, including Peachjar, to recruit 29 participants from their community for a longitudinal study of teen mental health. In this case study, we present our experience using Peachjar to recruit minor teen participants for an online behavioral experiment designed to assess how age and other demographic factors affect participants' ability to recognize giveaway scams found on YouTube. We found that Peachjar requires careful survey design to prevent spam and may be more expensive than other comparable methods. Our study received $55$ valid, complete responses after we began using Peachjar, with an estimated cost per participant of $\$43.18$.


%% file: secs/02_methods.tex
\section{Experimental Context}
\label{sec:methods}

In this section, we briefly describe our experimental design, recruitment procedure, and ethical considerations in our research methods. 

\subsection{Experimental Design}

Our experiment was designed to evaluate how age affects one's ability to identify YouTube scams. In particular, we presented participants with giveaway scams that falsely purport to show websites where one can get free electronic goods, such as in-game currency or subscriptions to online services. Previous research has speculated that these scams may target children~\cite{boumasims2021first}. We hypothesized that age would be inversely related to the rate of scam victimization, with minor teenagers (aged 13 to 17 years old) falling for the scams at a higher rate than adults. We also expected that other demographic and experimental factors (e.g., income level, past victimization by scams, etc.) may also relate to the victimization rate.

To assess these hypotheses, we developed an online scenario-based experiment in which participants were asked to advise a hypothetical friend searching on YouTube for ``free Spotify premium'' or ``free Roblox Robux.''\footnote{Robux are the in-game currency for the online game platform, Roblox} The experiment was designed to simulate the sequence of steps that may lead someone to be victimized by one of these scams. Participants were first presented with nine search results corresponding to their search terms and asked to select which (if any) they thought their friend should click on. 
Participants were shown two videos, one that presented a scam and the other that presented a legitimate free offer (e.g., a free trial). They were asked to advise their friend on what they should do next (e.g., visit the website, exit the video, report it to YouTube, etc.). Participants were also asked to review a video of someone scrolling through the website linked by the video and describe their recommended action. All tasks included follow-up questions that asked participants to explain their reasoning. Participants were also asked questions to assess potential confounding variables (e.g., demographics, previous victimization by scams, etc.).

\subsection{Recruitment Procedure}

We recruited two populations: a sample of adults and a sample of teenagers. For our sample of adults, we recruited an age-balanced sample of crowd workers from Prolific\footnote{\url{https://www.prolific.com/}} ($n = 205$). In the remainder of the paper, we focus on the teenage sample.

Teenagers were recruited via their parents/legal guardians through a digital flier (see Figure~\ref{fig:poster} in Appendix~\ref{app:recruitment}) that informed parents/guardians about the study and included a link to an informed consent form. To avoid priming participants about the nature of the study, we used nonspecific language to describe study tasks (i.e., "[we are] investigating how teens interact with social media posts on YouTube"). If a parent/guardian consented to their child participating in the study, they were redirected to a survey asking a few questions about their child's demographics (e.g., age, household income, race, etc.). Parents were then given a link to an assent form for their child to complete. 

The child assent form explained the study procedure and the potential risks of participation at an 8th-grade reading level (based on Flesch reading ease). Participants were required to affirm that they were eligible for the study (i.e., they were fluent in English, located in the US, used YouTube, and were between 13 and 17 years old). We also required participants to correctly identify a word spoken in a YouTube video to verify that they could access the media presented in the experiment. 

After the assent form, participants were redirected to the primary survey. After completing this survey, they were redirected to a final survey where they could enter their email address to be compensated with a \$5 Amazon gift card. We deleted the email addresses at the conclusion of the study.

The recruitment flier was initially distributed using snowball sampling~\cite{parker2019snowball}. It was shared with the parents of eligible teenage participants already known to the authors. These parents were asked to share the flier with other eligible teens' parents. Using this technique, we recruited 30 teenage participants from November 11, 2023, to December 7, 2023. After we stopped receiving responses from our initial sampling efforts, we were motivated to use Peachjar to recruit a larger and more diverse sample. We launched our first Peachjar fliers on December 13, 2023.

\subsection{Ethics}

Our study protocol was approved by our Institutional Review Board (IRB). The main ethical concern in developing our recruitment procedure was ensuring that teenagers received parental consent before taking the study. If we recruited teenagers directly, it was likely that many would pretend to be their parents when asked to obtain parental consent. In consultation with our IRB, we opted to recruit parents first and have them share the survey with their children. This procedure ensured that we received parental consent before the survey was seen by minors.  

As our survey included examples of actual fraud, we also took steps to protect participants from victimization. We redacted all URLs in videos presented to participants. At the end of the survey, we also provided a debrief identifying the scam videos participants saw and providing instructions on avoiding fraud. 

%% file: secs/03_results.tex
\section{Recruitment with Peachjar}
\label{sec:results}

In this section, we describe our recruitment on Peachjar in detail. We first discuss the characteristics of Peachjar before explaining how we distributed fliers and protected our survey from spam. We conclude by describing the results of our recruitment and the demographics of our teenage participants.

\subsection{Characteristics of Peachjar}
As of May 16th, 2024, $780$ school districts with $14,578$ schools were using Peachjar. The school districts are mostly public, although some private schools are included. The largest number of participating school districts are in California ($21.7\%$), Texas ($10\%$), and Washington ($7.6\%$) (see Table~\ref{tab:state_counts} in appendix~\ref{app:stats} for a count of school districts by state). 
The median number of schools per school district is eight elementary schools, two middle schools, and two high schools (see a plot of the distribution of school counts in Figure~\ref{fig:school_counts} in Appendix~\ref{app:stats}).

The cost of submitting a flier to a school is denominated in credits, with five credits required to submit a flier to a school. Credits are available at various prices, with discounts for first-time purchasers and those buying in bulk. The base cost per flier is \$25 (see Table~\ref{tab:cost_per_credit} in Appendix~\ref{app:stats} for a detailed breakdown of the cost per credit). Schools are not obligated to accept fliers; the credits are refunded when a school rejects a flier.  Additionally, some school districts ($9.4\%$) require fliers to be available in one or more languages other than English, most often Spanish.

\subsection{School Sampling}

Since it was financially infeasible for us to distribute fliers to all schools on Peachjar, we adopted various sampling techniques over time. We initially selected districts based on the median income where the school district was located. Median income was determined based on the 2022 American Community Survey (ACS) at the billing zip code provided for the school by Peachjar~\cite{acs2022}.\footnote{We used the field S1901\_C01\_012E, which is median household income over the last 12 months in 2022 inflation-adjusted dollars} This analysis is imperfect, as many students will not live in the zip code where the school district offices are located. Nevertheless, this gives us some insight into the socioeconomic status of the students in each district (see Figure~\ref{fig:income} in Appendix~\ref{app:stats} for a plot of the distribution of median income by school district).

We selected the largest school in school districts with at least one high school evenly from three income groups: less than \$42,606 ($23$ or $3.0\%$), \$42,606 to \$74,580 ($291$ or $37.3\%$), and greater than \$74,580 ($316$ or $40.5\%$). \$42,606 was the threshold for a three-person household to receive reduced-price lunches starting July 2022~\cite{USDAChild}. \$74,580 is the national median household income reported in the 2022 ACS~\cite{acs2022}. We used this method to select eight schools on December 13, 2023. We selected nine additional schools on 
January 3, 2024. Due to spam, we had to remove all fliers on January 4, 2024 (see section~\ref{ssec:security}). We relaunched our survey with 18 new schools on January 8, 2024.  

On February 26, 2024, we submitted our fliers to 50 more schools. We decided to select school districts by sampling those with the most high schools. We expected these may have the largest student populations and would result in more participants. We also translated our materials for parents (i.e., flier, consent form, and parental survey) into Spanish to send fliers to schools that required Spanish and English fliers. Teenage participants were still required to be fluent in English to take the survey. Five additional fliers were submitted to schools on April 30, bringing the total number of schools to 90.

Throughout the flier submission process, many schools rejected our flier or failed to respond to our request within Peachjar's time limit of 8 days from the intended distribution date. After rejection or time-out, we submitted requests to a new school. To place fliers at 90 schools, we had to submit requests to 176 school districts. While most fliers were accepted on or around the submission date, this led to a ``long tail'' of flier distribution after the above dates. For example, the last flier from the January 8, 2024 batch was not accepted until February 12. The last flier from the February 26 batch was not accepted until April 3. Most schools provided no feedback when rejecting our flier, but the most common reason provided was that it represented a form of solicitation. 

\subsection{Survey Security}
\label{ssec:security}
We initially did not implement security measures to prevent spam. We expected only parents in our selected school districts to see the fliers. However, Peachjar fliers have no access control, and anyone can view them. From January 2 to January 4, 2024, we received many responses that appeared to be spam (i.e., duplicate responses, meaningless answers to open-ended questions, failed Captcha, etc.). Once we noted this issue, we paused data collection and retracted all of our Peachjar fliers posted up to this point. Out of caution, we removed all survey responses (122) received during this period, including a small number that wrote plausible responses to open-ended questions. These participants were paid, but their data were excluded from the analysis.

We relaunched our survey on January 8, 2024 with new anti-spam checks. Rather than using the same fliers for all school districts, we generated fliers with a unique URL for each school district. Parents were now required to select the school district their child attended from a drop-down list to progress through the survey. Parents who selected a school district or state different from the one corresponding to their flier could not continue the study. 

This security setup was effective. Most attempted spammers were screened out by the requirement to identify the school district and state of the flier correctly. Moreover, using a custom URL for each flier allowed us to quickly block spam from a particular flier without shutting down the entire survey. We had to remove a flier on March 8, 2024, due to one or more spammers discovering the correct answers to the security questions. 125 responses from this flier on March 7 and 8, 2024 were removed. 

\input{tabs/demographics}

\subsection{Results of Peachjar Recruitment}

$55$ participants were recruited after we began using Peachjar. We received $22$ valid responses between December 13 and December 31, 2023. Although most of these responses were likely from Peachjar, some may have come from previously distributed fliers. The remaining $33$ responses must have come from Peachjar, as we disabled our previously distributed survey links on January 4, 2024. The total cost for 90 fliers was $\$1625$. Our cost per participant was an average of \$39.94.\footnote{This calculation is based on only the 33 participants we recruited via the 73 fliers with unique links.} We estimate that submitting a flier for each school takes approximately 3 minutes. At $\$15$ per hour, this would bring the average cost to $\$43.18$. 

While comparison is difficult due to the lack of studies reporting cost and differences in methods, this cost per participant seems to be high compared to studies using social media recruitment (although it may be lower than the cost of using a participant recruitment service or paying to access a participant panel). For example, a 2020 review of mental health studies using social media for recruitment found that the median cost per participant was \$19.47~\cite{SANCHEZ2020152197}. 

Table~\ref{tab:demographics} shows the distribution of demographic characteristics among the 55 participants recruited after we began distributing Peachjar fliers. While not representative,  the sample obtained from Peachjar was reasonably diverse. Participants were drawn from 23 different US states and the District of Columbia.

Each survey in our recruitment pipeline had fewer responses than the previous one. After December 13, 2023 (i.e., considering all teen participants), we received 955 responses to the parental consent form. Of these respondents, 900 were directed to the parental survey, 40 were screened out for not meeting the eligibility requirements, and 15 did not consent. We received 831 responses to the parental survey, of which 420 were screened out for failing the security check, and 411 were given a link to the child assent form. We received 367 responses to the child assent form. Of these respondents, 8 did not pass the audio check, 2 were screened out for being ineligible, 1 did not consent, and 356 were passed on to the main child survey. We received 320 responses to the child survey. As discussed in section~\ref{ssec:security}, 247 of these responses were removed as spam (122 responses from January 2 to 4, 125 responses from March 7 and 8). 18 additional responses were incomplete. The median completion time for the 55 valid and complete responses was 19 minutes and 44 seconds.

Most fliers resulted in no participants. Of the 73 fliers posted after we began using unique links in each flier, only $19$ resulted in at least one valid participant. $10$ of the $19$ fliers resulted in only one participant each, with five participants coming from the most successful flier. Interestingly, our income-based sampling technique seems to have resulted in more participants than targeting the largest school districts. The 55 fliers posted to the largest school districts resulted in only 16 valid responses,  compared to 17 valid responses from the 18 fliers posted on January 8, 2024.

%% file: tabs/demographics.tex
\begin{table*}[ht]
    \centering
    \caption{Demographics for the 55 participants recruited after we began using Peachjar. Multiple options could be selected for race/ethnicity}
    \label{tab:demographics}
    \begin{tabular}{cccccccc}\toprule
    \multicolumn{2}{c}{Gender} & \multicolumn{2}{c}{Age} & \multicolumn{2}{c}{Race/Ethnicity} & \multicolumn{2}{c}{Household Income} \\
    \cmidrule(lr){1-2}\cmidrule(lr){3-4}\cmidrule(lr){5-6}\cmidrule(lr){7-8}
    Boy & $51\%$ & 13 & $9\%$ & 
American Indian or Alaska Native & $2\%$ & Less than \$20,000 & $6\%$ \\
    Girl & $47\%$ & 14 & $20\%$ & Asian & $11\%$ & \$20,000 to \$39,999 &  $15\%$\\
    Non-binary & $0\%$ & 15 & $38\%$ &  Black or African American & $16\%$ & \$40,000 to \$59,999 & $18\%$\\
    Self-describe & $0\%$ & 16 & $18\%$ & Hispanic and/or Latino/a/x & $7\%$ & \$60,000 to \$79,999 & $7\%$ \\
    No response & $2\%$ & 17 & $15\%$ & White & $73\%$ & \$80,000 to \$99,999 & $9\%$\\
    & & & & Self describe & $0\%$ & \$100,000 to \$149,999 & $20\%$ \\
    & & & & No response & $7\%$ & Over \$150,000 & $11\%$ \\
    & & & & & & No Response & $14\%$\\
    \bottomrule
    \end{tabular}
\end{table*}

%% file: secs/04_discussion.tex
\section{Conclusion}
\label{sec:discussion}
We conclude that Peachjar can be an effective way to recruit a geographically diverse sample of teenage participants for online studies, but it requires a lot of effort to screen out spam and the number of successful recruits per school is extremely low. The cost per participant may also be higher than for other methods of online recruitment.  
Although the requirement that participants view videos may have harmed our conversion rate slightly, the main challenge seems to be getting parents interested in filling out the initial survey. It is unclear whether a higher compensation rate would have been helpful or if there are different design choices that we could have implemented to make our flier more compelling. We did not try resubmitting fliers to schools with higher conversion rates or submitting fliers to other schools in those districts; these approaches might be worth trying. Compensating parents for participation (for example, by entering them in a giveaway for a gift card) may have helped motivate participation. Any compensation scheme for parents must be carefully crafted to avoid coercion. In addition, using Peachjar for recruitment requires a lot of lead time, as many schools using the service never responded to our requests to publish fliers, leading to long delays before all recruitment fliers could be posted.

%% file: secs/05_app.tex
\clearpage
\appendix
\onecolumn
\section{Recruitment flier}
\label{app:recruitment}
\begin{figure*}[htp]
    \centering
    \includegraphics[width=0.55\textwidth]{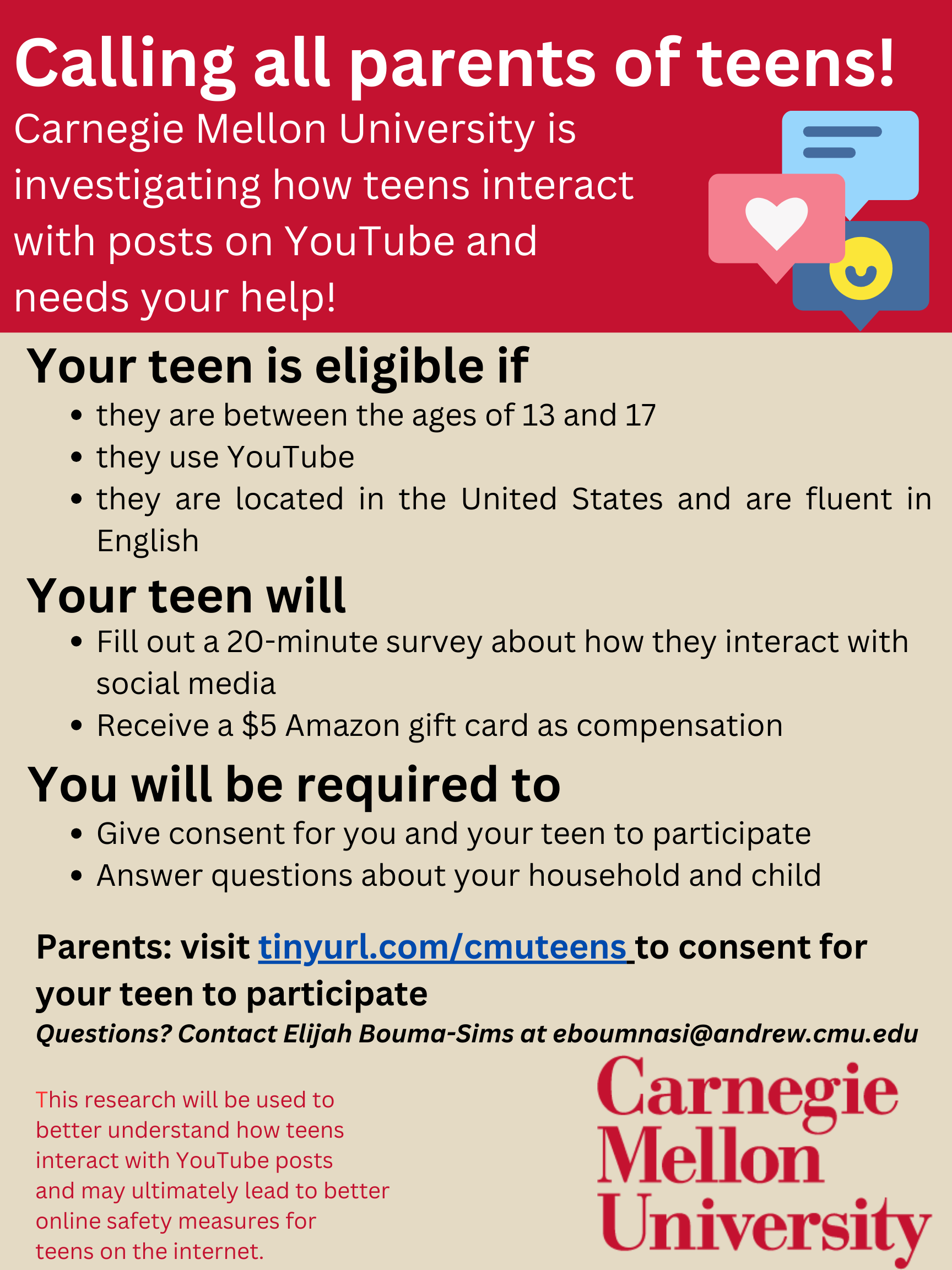}
    \caption{Poster used to recruit parents of teenage participants.}
    \label{fig:poster}
\end{figure*}
\newpage
\section{Additional Peachjar Statistics}
\label{app:stats}
\begin{table*}[htp!]
\centering
\caption{Count of school districts and schools using Peachjar by US State as of May, 2024}
\label{tab:state_counts}
\begin{tabular}{ccc|ccc}
\toprule
\textbf{State} & \textbf{\# Districts} & \textbf{\# Schools} & \textbf{State} & \textbf{\# Districts} & \textbf{\# Schools} \\ 
\midrule
AL &   5 &  52 & NC &  26 & 760 \\  
AR &   6 & 115 & ND &   1 &  16 \\ 
AZ &  33 & 574 & NE &   3 &  27 \\ 
CA & 169 & 3346 & NJ &   2 &  20 \\
CO &  13 & 500 & NM &  10 & 298 \\
CT &  2 &  40 & NV &   1 &   9 \\ 
DE &   3 &  49 & NY &  17 & 108 \\ 
FL &  17 & 738 & OH &  26 & 327 \\ 
GA &  12 & 357 & OK &  14 & 271 \\ 
IA &   5 &  66 & OR &  20 & 247 \\ 
ID &   9 & 116 & PA &  19 & 141 \\ 
IL &  14 & 171 & SC &  19 & 361 \\ 
IN &  11 & 242 & SD &   3 &  49 \\
KS &   9 & 162 & TN &   6 &  80 \\
KY &   1 &   3 & TX &  78 & 1728 \\
LA &   1 &  33 & UT &   7 & 189 \\ 
MD &   7 & 216 & VA &  13 & 339 \\ 
MI &  25 & 261 & WA &  59 & 1153 \\
MN &  37 & 412  & WI &  18 & 180 \\ 
MO &  50 & 707 & WV &   1 &  25 \\ 
MS &   7 &  80 & WY &   1 &  10 \\
\bottomrule
\end{tabular}
\end{table*}

\begin{figure}
    \centering
    \includegraphics[width=0.75\paperwidth]{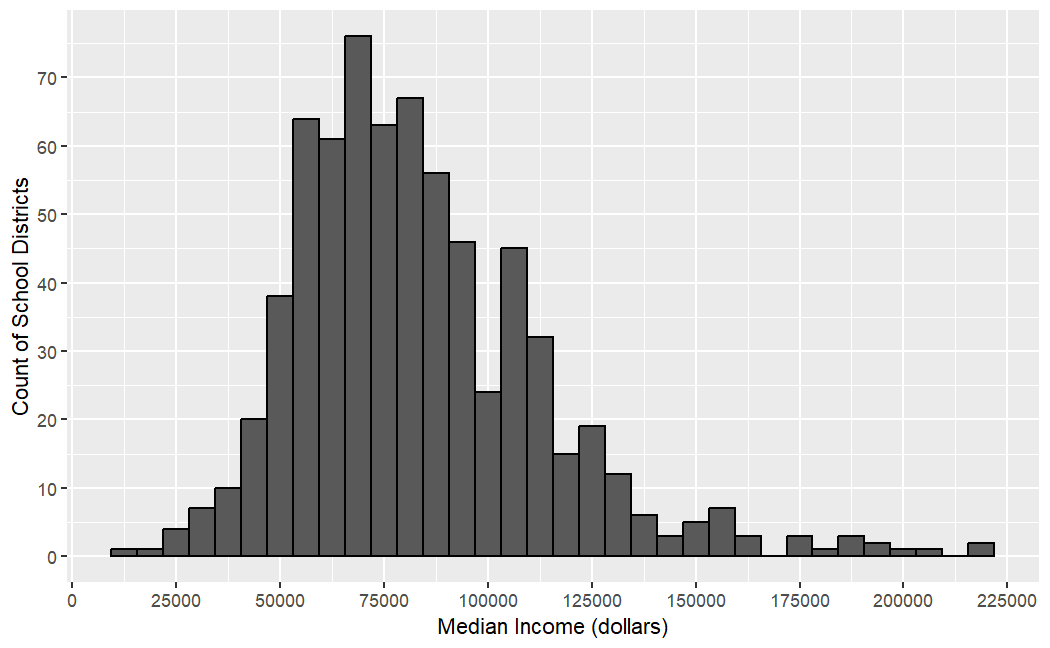}
    \caption{Distribution of median household 2022 income at Peachjar billing zip codes as of May 2024. 82 school districts located in zip codes where no ACS data was available were excluded. An older list of schools from November 2023 was used for sampling.}
    \label{fig:income}
\end{figure}

\begin{figure}
    \centering
    \includegraphics[width=0.75\paperwidth]{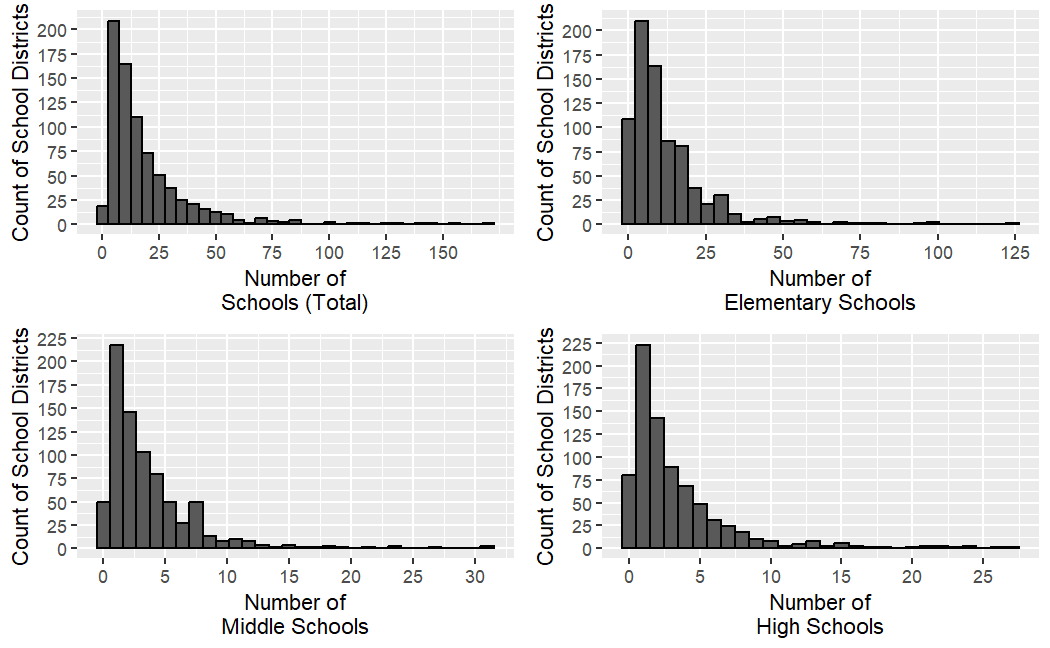}
    \caption{Distribution of number of schools in each school district as of May 2024.  Categorizations are provided by Peachjar}
    \label{fig:school_counts}
\end{figure}

\input{tabs/cost_per_credit}

%% file: tabs/cost_per_credit.tex
\begin{table*}[htp]
    \centering
    \caption{Cost of distributing a flier per credit/school district as of February 2024. To distribute 90 fliers, we first purchased a ``starter pack'' of $200$ credits for $50\%$ off ($\$500$). We then purchased $250$ credits for $\$1125$. Peachjar indicated that those purchasing $2500$ or more credits ($500$ schools) may receive a better, unspecified rate.}
    \label{tab:cost_per_credit}
    \begin{tabular}{cccccc}
    \toprule
    \# of School districts & Discount & Price/School & Total Price & \# of Credits & Price per Credit \\
    \midrule
    1 & 0\% & \$25.00 & \$25 & 5 & \$5.00 \\
    20 & 10\% & \$22.50 & \$450 & 100 & \$4.50 \\ 
    60 & 15\% & \$21.25 & \$1275 & 300 & \$4.25 \\
    150 & 20\% & \$20.00 & \$3000 & 750 & \$4.00 \\
    \bottomrule
    \end{tabular}
\end{table*}